\documentclass[10pt, conference]{IEEEtran}
\IEEEoverridecommandlockouts

\makeatletter
\newcommand{\newlineauthors}{%
  \end{@IEEEauthorhalign}\hfill\mbox{}\par
  \mbox{}\hfill\begin{@IEEEauthorhalign}
}
\makeatother

\usepackage{float}
\usepackage[style=ieee]{biblatex}
\usepackage{amsmath}
\usepackage{color,soul} % to highlight text with \hl command
\usepackage[colorlinks=true, allcolors=blue]{hyperref}
\usepackage[nocomma]{optidef}
\usepackage{amssymb}
\usepackage{bm}
\usepackage{subcaption}
\usepackage{placeins}
\usepackage{lipsum}

\newcommand{\aFRR}{\text{aFRR}}
\newcommand{\mFRR}{\text{mFRR}}

\newcommand{\B}{\mathrm{B}}
\newcommand{\BRP}{\mathrm{BRP}}

\begin{document}

\title{Herd Behavior in Decentralized Balancing Models: A Case Study in Belgium
% \thanks{979-8-3195-3554-2/26/\$31.00 ©2026 IEEE}
}

% Author list: 
\author{\IEEEauthorblockN{Max Bruninx}
\IEEEauthorblockA{\textit{Vrije Universiteit Brussel}\\
Brussels, Belgium\\
  max.bruninx@vub.be
  }
\and
\IEEEauthorblockN{Seyed Soroush Karimi Madahi}
\IEEEauthorblockA{\textit{Ghent University}\\
Ghent, Belgium\\
seyedsoroush.karimimadahi@ugent.be
}
\and
\IEEEauthorblockN{Timothy Verstraeten}
\IEEEauthorblockA{\textit{Vrije Universiteit Brussel}\\
Brussels, Belgium\\
timothy.verstraeten@vub.be
}
\newlineauthors
\IEEEauthorblockN{Jan Decuyper}
\IEEEauthorblockA{\textit{Vrije Universiteit Brussel}\\
Brussels, Belgium\\
jan.decuyper@vub.be
}
\and
\IEEEauthorblockN{Chris Develder}
\IEEEauthorblockA{\textit{Ghent University}\\
Ghent, Belgium\\
chris.develder@ugent.be
}
\and
\IEEEauthorblockN{Jan Helsen}
\IEEEauthorblockA{\textit{Vrije Universiteit Brussel}\\
Brussels, Belgium\\
  jan.helsen@vub.be
  }
}

\maketitle

\begin{abstract}
    In a decentralized balancing model, Balance Responsible Parties (BRPs) are encouraged by the Transmission System Operator (TSO) to deviate from their schedule to help the system restore balance, also referred to as implicit balancing. This could reduce balancing costs for the grid operator and lower the entry barrier for flexible assets compared to explicit balancing services. However, these implicit reactions may overshoot when their total capacity is high, potentially requiring more explicit activations. This study analyses the effect of increased participation in the decentralized balancing model in Belgium. To this end, we develop a market simulator that produces price signals on minute-level and simulate the implicit reactions for battery assets with different risk profiles. Besides the current price formula, we also study two potential candidates for the near-term presented by the TSO. A simulation study is conducted using Belgian market data for the year 2023. The findings indicate that, while having a significant positive effect on the balancing costs at first, the risk of overshoots can outweigh the potential benefits when the total capacity of the implicit reactions becomes too large. Furthermore, even when the balancing costs start to increase for the TSO, BRPs were still found to benefit from implicit balancing.
\end{abstract}


\section{Introduction}

Global renewable energy capacity continues to grow sharply year-on-year \cite{IEA_GlobalEnergyReview2025}. While this helps to reduce fossil fuel consumption and global emissions, their intermittent nature poses challenges for the grid to remain in balance. Therefore, Transmission System Operators (TSOs) need to rely on the activation of grid reserves to ensure stability of the power system. The costs related to these services are attributed to Balance Responsible Parties (BRPs), who are responsible for keeping their portfolio of generators and off-takers balanced, through the imbalance settlement procedure.

The main objective of imbalance settlement is to incentivize BRPs to maintain balance within their portfolio by engaging in day-ahead and intraday trading. Any deviations from their contracted schedule are therefore penalized with the imbalance price. However, since single imbalance pricing became prevalent in Europe \cite{acer2020harmon}, BRPs can also be rewarded during imbalance settlement when their open positions contribute to restoring balance in the system. In a decentralized balancing model, the TSO even encourages BRPs to deviate from their schedule to help balance the grid in real-time \cite{elia2025flexibility}. We will further refer to this deliberate deviation from the contracted volumes as \textit{implicit balancing}. To incentivize BRPs, the TSO publishes an (intermediate) real-time price signal that should guide their reactions. This model can reduce grid balancing costs and lower the entry barrier for flexible assets compared to frequency restoration reserves, though it requires the TSO to sacrifice explicit control of balancing actions. An important drawback of the decentralized balancing model is the introduction of short-term oscillations in the system, which occur when the implicit reaction is larger than the required volume to restore balance. With the growing share of flexible assets, such as grid-scale batteries or curtailable renewables, this issue will likely become more pronounced. Therefore, it is essential that the TSO carefully establishes real-time price signals that align participant behavior with system balancing needs.

Past research on implicit balancing focused on developing control strategies that could increase operational profit for BRPs. To incorporate the impact of these strategies on the imbalance price, methods were proposed employing model predictive control \cite{smets2023strategic, wessel2024risk}, reinforcement learning \cite{rasic2025safe} or a combination of both \cite{madahi2025model}. However, apart from \cite{madahi2025model}, they only consider a granularity of 15 minutes, whereas intermediate prices are published every minute in Belgium and the inter-quarter hour dynamics are also important to account for \cite{soroush2025gaming}. Furthermore, they focus on one market participant aiming to optimize its position using state-of-the-art forecast and optimization methods, rather than all market participants acting upon the signal provided by the TSO.

In this work, we want to take a more holistic view and study the impact of a growing number of flexible assets which participate in implicit balancing of the Belgian grid. To achieve this, we develop a market simulation model which outputs minute-level price signals and incorporate a feedback loop that simulates additional implicit response to these signals. To determine the explicit activations by the TSO, the simulator applies a mixed integer linear program to historical data of the individual balancing energy bids. The model produces price signals for every timestep by considering the imbalance price formula and the information on the activated reserves during the imbalance settlement period (ISP). In addition to the current formula, we consider two formulas proposed for the near future by the Belgian TSO (Elia) in their ``Real-time price design note I'' \cite{Elia2025RealtimePriceII}. The implicit response to those signals is modelled as a collection of battery energy storage system (BESS) assets with varying properties and risk appetites. These assets are assumed to respond to the price signals provided by the TSO rather than position themselves based on price forecasts.

The main contributions of the paper are:
\begin{itemize}
    \item We develop a market simulation model, extending previous work of \cite{bruninx2025day,soroush2025gaming}, which incorporates the feedback loop between the actions of the TSO and the reactions of BRPs. This simulator allows to study implicit balancing from the perspective of both parties for different scenarios.
    \item This paper compares three different imbalance price formulas, the current one as well as two potential candidates for the near future \cite{Elia2025RealtimePriceII}.
    \item This paper illustrates how the projected expansion of flexible capacity \cite{EliaAdequacyFlexibility2025} could disturb grid stability when BRPs do not account for the impact of their reaction on the market.
\end{itemize}

The simulator is applied to Belgian market data of 2023. We find that a greater implicit response to the imbalance price can have a positive impact on the balancing costs, while also being profitable for BRPs. However, when the total capacity becomes too large, the risk of overshoots outweighs the potential benefits of implicit balancing. To resolve this issue, the TSO could incentivize BRPs to pursue more advanced control methods accounting for their market impact as well as the reactions of other parties.

\section{Market simulation}
\label{sec:market_simulator}

\begin{figure}[t!]
    \centering
    \includegraphics[width=0.8\linewidth]{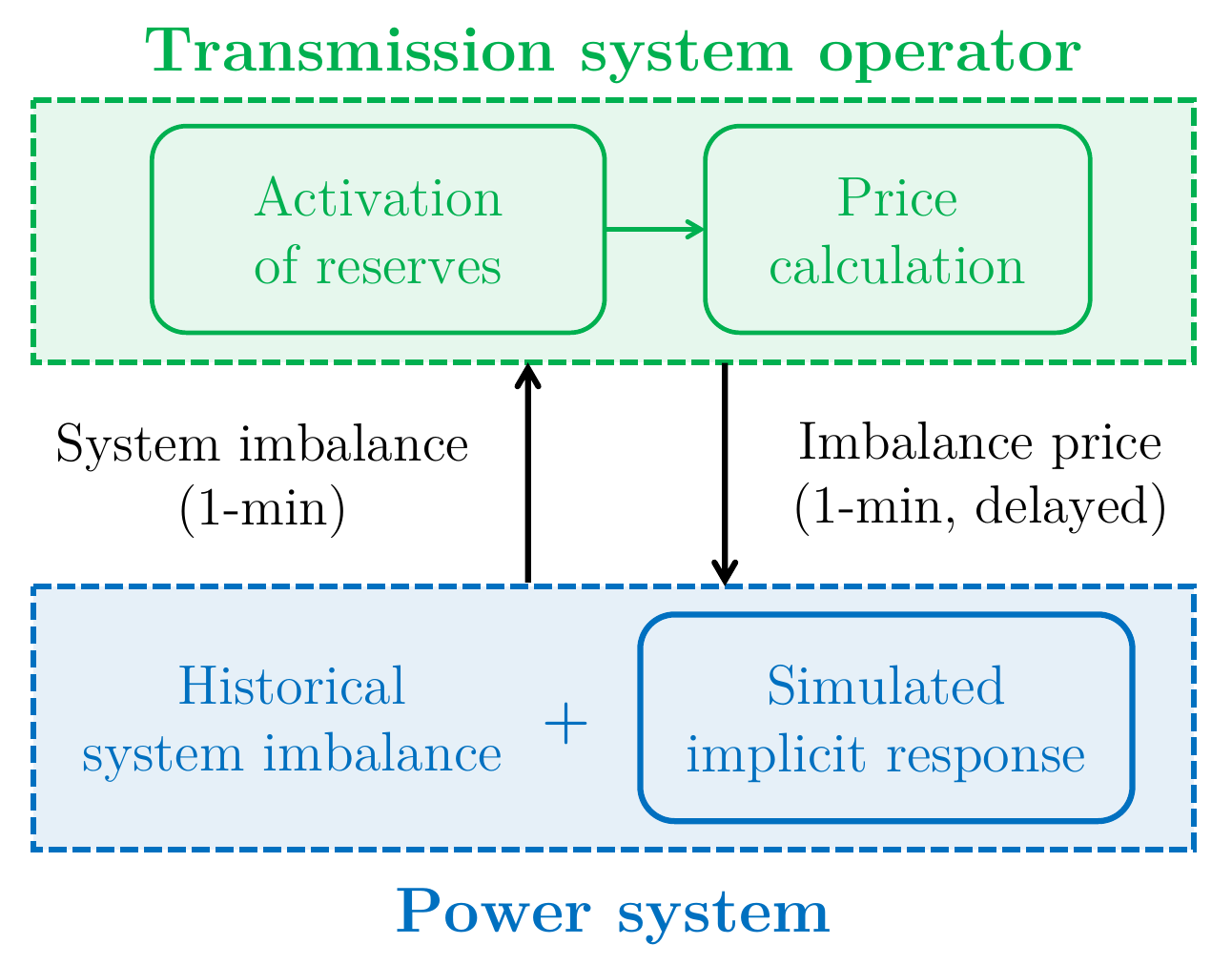}
    \caption{High-level overview of the simulation environment modelling the feedback loop between the power system and the transmission system operator. Every minute, the TSO observes the actual system imbalance, whereas the power system only receives the imbalance price with a certain delay (e.g., 2 minutes in Belgium).}
    \label{fig:overview}
\end{figure}

An accurate analysis of implicit balancing requires accounting for the feedback loop between the imbalance prices published by the TSO and the positions adopted by BRPs. Specifically, BRPs respond to real-time price signals, which modifies the system imbalance and the subsequent activation of reserves. This change in the activated reserves will, in turn, alter the imbalance price and trigger a new response from BRPs. Figure \ref{fig:overview} provides a high-level overview of the simulation environment employed to model this feedback loop. The upper part (green) contains the two required steps to obtain the (intermediate) imbalance price given the evolution of the system imbalance. For every timestep, the TSO activates the required capacity of automatic and manual Frequency Restoration Reserves (aFRR and mFRR) which resolve the system imbalance at minimal cost (Section \ref{sec:market_clearing}).  The 1-minute imbalance price is afterwards calculated based on all activations during the imbalance settlement period up to that timestep (Section \ref{sec:price_calculation}). The lower part (blue) determines the system imbalance based on a combination of the historical system imbalance as well as a simulated implicit response to the published (intermediate) prices (Section \ref{sec:sim_implicit_response}).

\subsection{Activation of reserves}
\label{sec:market_clearing}

The TSO aims to activate the reserves which restore system balance at minimal activation cost. For each minute $t$, the corresponding optimization problem is given by:
\begin{mini!}|s|
    {}{
        \sum_{a^+\in \aFRR^+}\mu^{a^+}b^{a^+}_t
        - \sum_{a^-\in \aFRR^-}\mu^{a^-}b^{a^-}_t
    }{\label{opti-obj}}{}%\notag
    \breakObjective{
        + \sum_{m^+\in \mFRR^+}\mu^{m^+}b^{m^+}_t
        - \sum_{m^-\in \mFRR^-}\mu^{m^-}b^{m^-}_t,
    }\notag
    % part 1: required activations
    % def aFRR variables
     \addConstraint{\sum_{a^{+/-}\in \aFRR^{+/-}}b^{a^{+/-}}_t}{= V^{\aFRR^{+/-}}_t: \lambda^{\aFRR_{+/-}}_t,}{\label{opti-con:required_afrr_up}} 
     % def mFRR variables
    \addConstraint{\sum_{m^{+/-}\in \mFRR^{+/-}}b^{m^{+/-}}_t}{= V^{\mFRR^{+/-}}_t:\lambda^{\mFRR^{+/-}}_t,}{}
    % resolve system imbalance
    \addConstraint{V^\aFRR_t=}{\ V^{\aFRR^{+}}_t-V^{\aFRR^{-}}_t,}{}
    \addConstraint{V^\mFRR_t=}{\ V^{\mFRR^{+}}_t-V^{\mFRR^{-}}_t,}{}
    \addConstraint{-(SI_t + V^\BRP_t)=}{\ V^\aFRR_t+V^\mFRR_t,}{\label{opti-con:resolve_imbalance}}
    % \addConstraint{-(SI_t + V^\BRP_t)=}{\ \left(V^{\aFRR^{+}}_t-V^{\aFRR^{-}}_t\right)\break+\left(V^{\mFRR^{+}}_t-V^{\mFRR^{-}}_t\right),}{\label{opti-con:resolve_imbalance}}
    % variable domain
    \addConstraint{0\leq V^{\aFRR^{+/-}}_t}{\leq\overline{V}^{\aFRR^{+/-}}},{\label{opti-con:afrr_limit}}
    \addConstraint{0\leq V^{\mFRR^{+/-}}_t}{\leq z^{+/-}_t\ \overline{V}^{\mFRR^{+/-}},}{}
    \addConstraint{\mathclap{z_t^{+/-}=\begin{cases}
            1 \quad V_t^{\aFRR^{+/-}} = \overline{V}_t^{\aFRR^{+/-}}\\
            0 \quad \text{otherwise,}
        \end{cases}}}{}{\label{opti-con:activation_order}}
    % \addConstraint{0\leq b^r_t\leq B^r,}{\quad\forall r\in \mathcal{R},}{\label{opti-con:individual_capacity}} 
    \addConstraint{0\leq b^{r}_t\leq B^{r},}{\quad\forall r\in \mathcal{R},}{\label{opti-con:individual_capacity}} 
\end{mini!}
where $\mathcal{R}$ the set of all individual balancing bids, $SI_t$ the historical system imbalance per minute and $V^\BRP_t$ the total simulated response. The activated volume of an individual bid is represented by $b^{r}_t$ and the corresponding price per volume by $\mu^{r}_t$. For the individual bids, the reserve type is reflected by the superscript $a$ for aFRR and $m$ for mFRR. The superscript \scalebox{0.8}{$+/-$} defines the direction of the balancing energy, with \scalebox{0.8}{$+$} and \scalebox{0.8}{$-$} representing incremental and decremental energy, respectively. We use the symbol $\overline{\left(.\right)}$ to denote the total capacity of each reserve type and $B^r$ for the individual capacity of each bid. Constraints (\ref{opti-con:required_afrr_up})-(\ref{opti-con:resolve_imbalance}) ensure that sufficient aFRR and mFRR bids are activated to cover for the observed system imbalance. The activation order between aFRR and mFRR, together with their capacity limits, is retained using constraints (\ref{opti-con:afrr_limit})-(\ref{opti-con:activation_order}). Last, constraint (\ref{opti-con:individual_capacity}) limits the individual activations by their offered capacity.

Note that three simplifications compared to the real-world setting were made in the simulator. First, the Belgian TSO uses an aFRR optimization cycle of 4s instead of 1min. However, imbalance data is only made available on minute-level \cite{eliaopendata} and therefore this is the finest granularity we have access to. Second, our simulator does not consider the effect of cross-border sharing of reserves, currently in place due to the PICASSO \cite{entsoe_picasso} and MARI \cite{entsoe_mari} projects, since the individual balancing bids for neighboring countries are not publicly available. In our experiments we also use data of 2023, which is before the integration of Belgium in these two projects. Third, we are assuming a reactive balancing strategy\footnote{See \cite{haaberg2016classification} for a comparison between proactive and reactive balancing strategies.} and do not split the activation of reserves in two separate stages (like, for instance, \cite{allard2024forecast}). An interesting pursuit for future research would be to improve this part such that the balancing strategy is more representative of the actual strategy pursued by the TSO.

\subsection{Price calculation}
\label{sec:price_calculation}

To calculate the (intermediate) balancing price for a certain moment in time, the clearing information during the imbalance settlement period up to that moment is used. This study considers three different formulas to calculate the imbalance price: the currently adopted formula as well as two potential replacement candidates (``max/min with smoothed deadband'' and ``weighted average with dynamic weights'') \cite{Elia2025RealtimePriceII}. While each formula is different, they all rely on the same price components: the aFRR price $\lambda^\aFRR_t$, the mFRR price $\lambda^\mFRR_t$ and the spot price $\lambda^{\mathrm{spot}}_t$. In general, we can state that the imbalance price $\lambda^{\B}_t$ is given by:
\begin{equation}
    \lambda^{\mathrm{B}}_t = f\left(\lambda^{\aFRR}_t, \lambda^{\mFRR}_t, \lambda^{\mathrm{spot}}_t; \theta_t\right),
\end{equation}
where $\theta_t$ represents the parameters specific to the formula. In the remainder of this section, we will focus on two important differences between those formulas. The exact definition of the formulas and components can be found in Appendix \ref{sec:sim-price_components}. 

A first difference between the formulas is observed in the so-called deadband region\footnote{Following the same naming convention as used by Elia.}, i.e., the region for which the system imbalance is between $-25$ and $25$ MW. In the current imbalance price formula, the price in this region is constant and set equal to the spot price component. The reasoning behind this approach is that for such moderate system imbalances, the TSO does not want BRPs to react. However, when the system imbalance moves slightly outside of this region, the imbalance price can jump significantly. To resolve this issue, the Belgian TSO has proposed to introduce a smoothing function which dynamically weights the spot price and the other components based on the system imbalance. The corresponding imbalance price formula is referred to as ``max/min with smoothed deadband''.

A second difference occurs when mFRR bids are activated that are more expensive than the aFRR component $\lambda^\aFRR$. In the current as well as the ``max/min with smoothed deadband'' formula, the most expensive mFRR bid will drive the price. While this provides a strong price signal to resolve system imbalance, extreme prices can be observed even when activating a small volume of the most expensive bid during the ISP. The ``weighted average with dynamic weights'' formula, on the other hand, applies volume weighting of the aFRR and mFRR volumes which makes the prices less extreme. Furthermore, it has the added benefit of providing a more accurate price signal in case of overshoots or undershoots of the activated reserves \cite{Elia2025RealtimePriceII}. 

\subsection{Simulated implicit response}
\label{sec:sim_implicit_response}

We model the implicit response of BRPs to the price signals as different BESS assets which determine their set-points using bang-bang control with price thresholds:
\begin{equation}
    P_t = \begin{cases}
        \overline{P_t}\quad \text{if }\lambda^\B_t>\overline{\lambda}^B\\
        \underline{P_t}\quad \text{if }\lambda^\B_t<\underline{\lambda}^B\\
        0\quad\text{else,}
    \end{cases}
\end{equation}
where $\overline{P_t}$ is the maximum discharge power and $\underline{P_t}$ is the maximum charge power. The latest published imbalance price is given by $\lambda^\B_t$ and the price thresholds by $\overline{\lambda}^B$ and $\underline{\lambda}^B$. To model these assets in a realistic way, we also include state-of-charge and cycle limit constraints on the power output. Note that this control strategy directly relies on the price signals provided by the TSO and does not account for the price impact of the asset. The reason why we assume such a control strategy is because we want to study the expected behavior within a decentralized balancing model.

To consider the fact that market participants can have different risk preferences, the price thresholds are calibrated based on the trade-off between expected profit $\mathbb{E}\left[\pi\right]$ and conditional-value-at-risk $\text{CVaR}_{0.05}(\pi)$. In particular, we optimize the thresholds to find the minimum for:
\begin{equation}
    w\:\text{CVaR}_{0.05}\left(\pi\right) - (1-w)\:\mathbb{E}\left[\pi\right],
\end{equation}
where $\text{CVaR}_{0.05}(\pi)$ and $\mathbb{E}\left[\pi\right]$ are min-max normalized. The parameter $w$ is used to define different risk groups, namely, risk-averse ($w=0.8$), medium risk ($w=0.5$) and risk-neutral ($w=0$). The total simulated capacity is always distributed among these groups.

\section{Simulation study}
\label{sec:simulation_study}

\subsection{Set-up}

We perform a simulation study considering the year 2023, which was pointed out by Elia as ``the most recent year with normal conditions'' \cite{Elia2025RealtimePriceII}. The price thresholds are calibrated using the actual 1-min imbalance prices of 2022. For the total simulated capacity, we assume a C-rate\footnote{The C-rate is the ratio between the power and energy capacity of the battery.} of $0.5$. Moreover, this capacity is distributed among the different risk groups as follows: 20\% risk-neutral, 60\% medium risk and 20\% risk-averse. This is based on the assumption that most assets will try to balance risk and return. In Appendix \ref{sec:app_sensitivity_study}, a sensitivity study can be found to illustrate how the results would change when the total capacity is distributed differently. The different price formulas will also be referred to as the ``Current'' (\ref{eq:sim-current_price}), ``MMSD'' (\ref{eq:sim-prop_I}) and ``WADW'' (\ref{eq:sim-prop_II}) formulas.

\subsection{Results}
\label{sec:results}

\begin{figure}[ht!]
    \centering
    \includegraphics[width=0.86\linewidth]{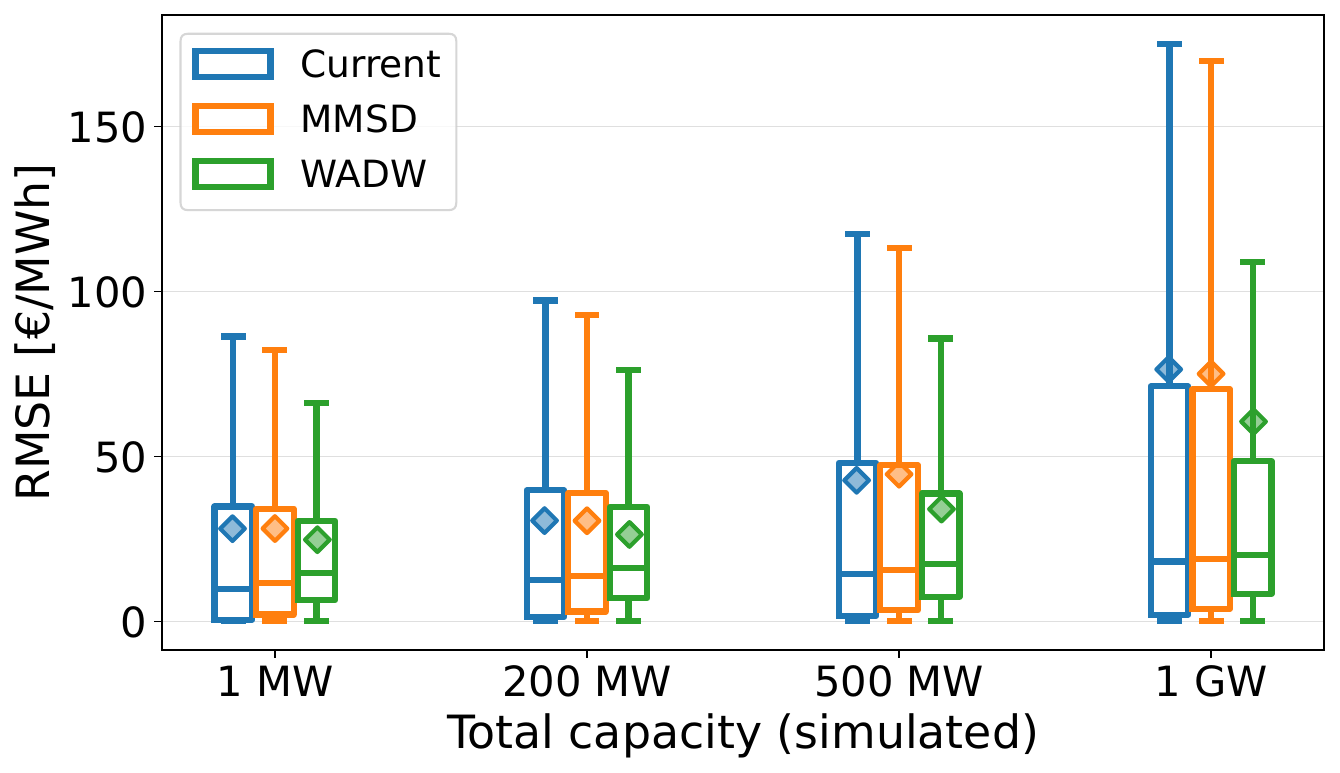}
    \caption{Boxplots of the RMSE of the minute-level imbalance price compared to the settlement price, with the diamond marker indicating the mean value. The RMSE is calculated for each imbalance settlement period seperately.}
    \label{fig:rmse}
\end{figure}

In Figure \ref{fig:rmse}, the root mean squared error (RMSE) per ISP of the intermediate prices compared to the settlement price are shown for each price formula. The measure indicates the accuracy of using the real-time price signal as an estimate for the actual settlement price. First, we observe a discrepancy between the mean and the median of the different price formulas. In particular, the median RMSE for the ``WADW'' formula is consistently higher than for the other formulas, whereas for the mean RMSE the opposite is true. This suggest that this formula is more robust to outliers (i.e., extreme prices), for which we have already provided a potential explanation in Section \ref{sec:price_calculation}. Second, we note that the spread (IQR and whisker range) is more moderate for this formula compared to the others. However, we do observe that the ``Current'' and ``MMSD'' formula attain more probability mass around zero. A reason could be that in these two formulas it is more likely that the price stays constant even when more reserves are activated (in the deadband region and when expensive mFRR bids are activated). Last, we find that the spread for the ``Current'' and ``MMSD'' formulas grows stronger when the simulated capacity increases. This is likely because of overshoots causing expensive bids to be activated which has a greater impact for these two formulas.

\begin{figure}[t!]
     \centering
     \begin{subfigure}[b]{0.42\textwidth}
         \centering
         \includegraphics[width=\textwidth]{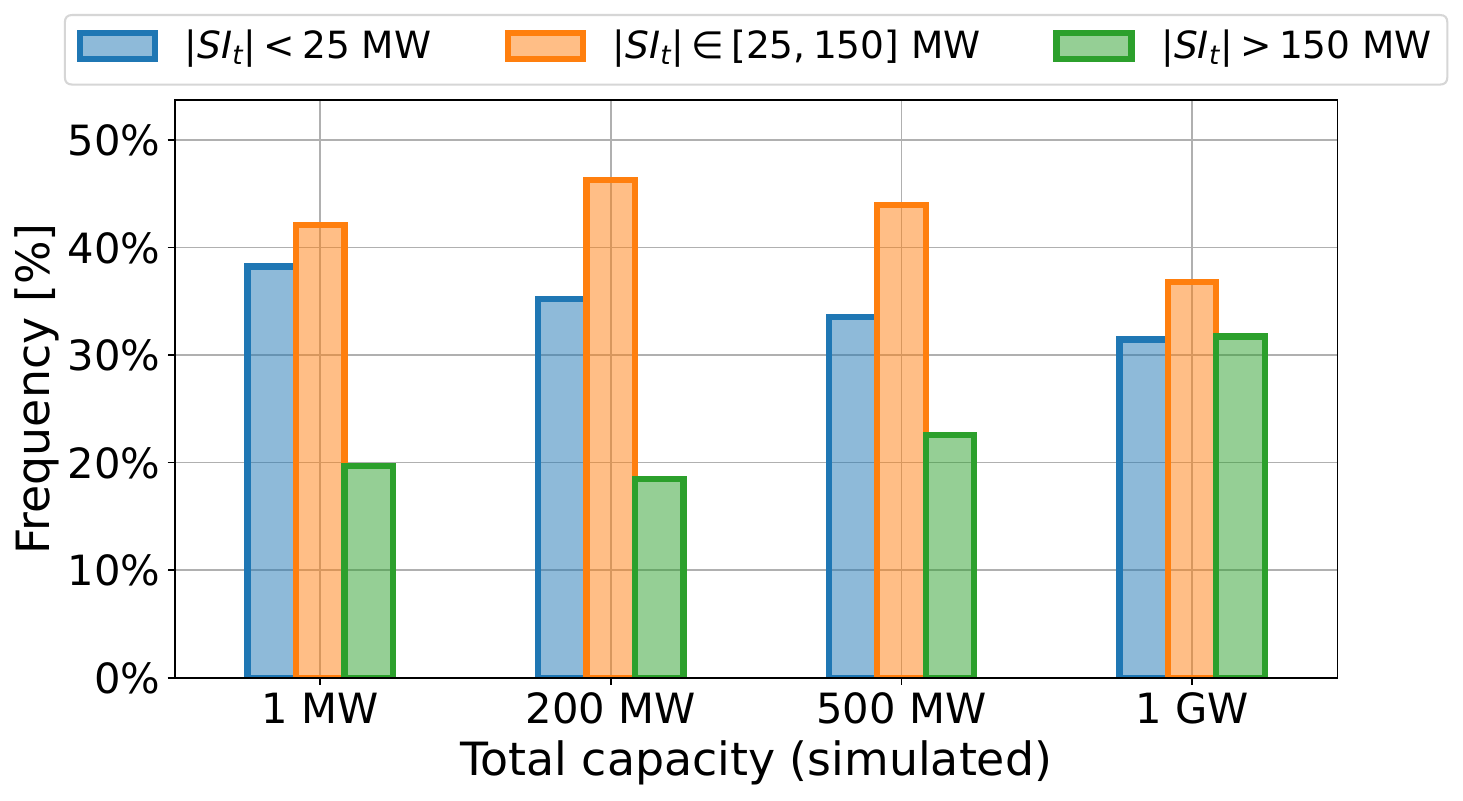}
         \caption{1-min}
         \label{fig:system_imb_1min}
     \end{subfigure}
     
     \vspace{12pt}
     
     \begin{subfigure}[b]{0.42\textwidth}
         \centering
         \includegraphics[width=\textwidth]{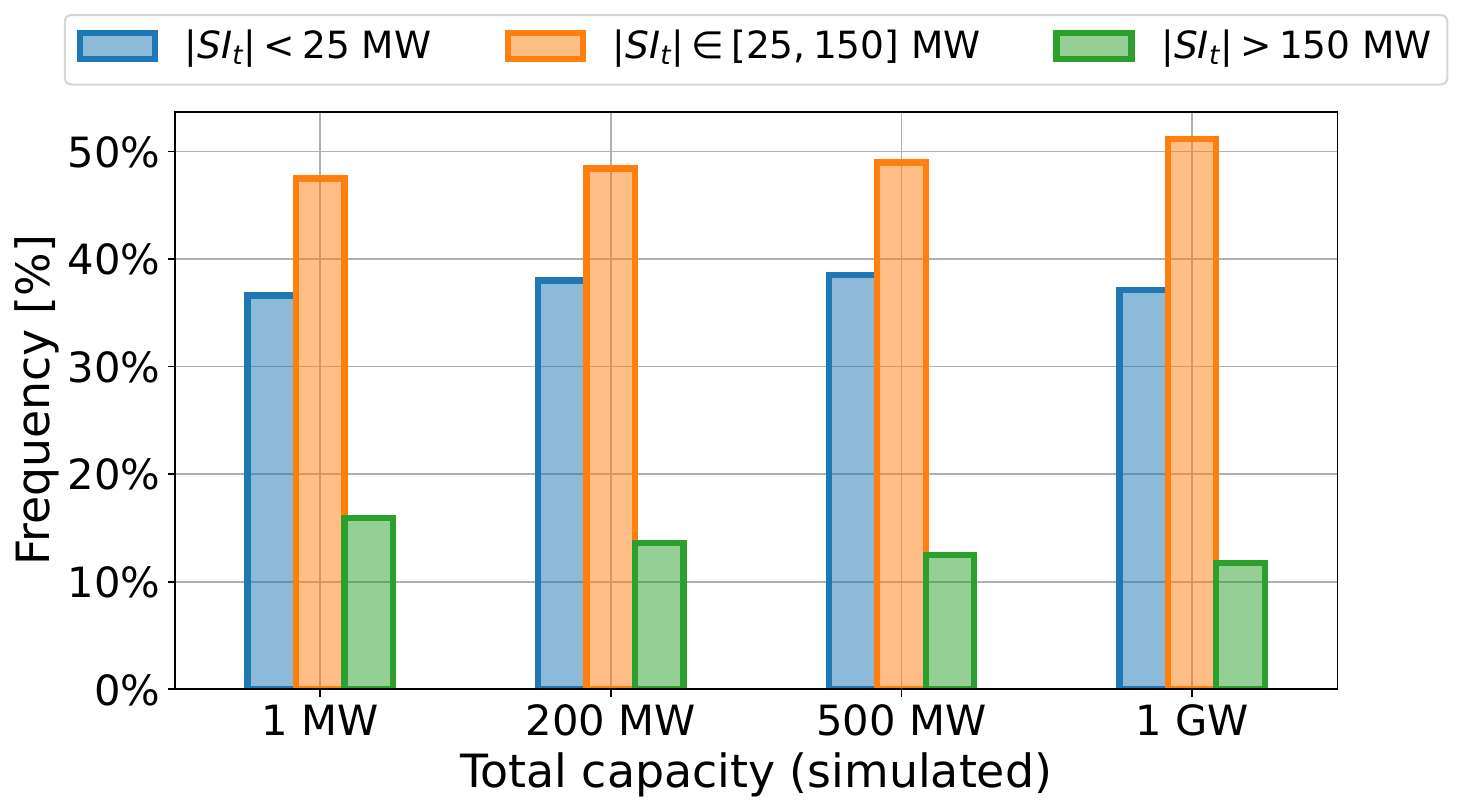}
         \caption{15-min}
         \label{fig:system_imb_15min}
     \end{subfigure}
     \caption{
     Distribution of the absolute system imbalance (1-minute and 15-minute level) using the current price formula. The imbalances are binned according to the three regions considered in the price formula: the deadband ($<$$25$ MW), normal system imbalance ($25$$-$$150$ MW) and extreme system imbalance ($>$$150$ MW). The calculations were made based on the simulations for the current price formula.
     }
     \label{fig:system_imbalance}
\end{figure}

Figure \ref{fig:system_imbalance} displays the difference between the distribution of the 1-minute system imbalance and the average system imbalance over the imbalance settlement period (15-minute). The absolute values of the system imbalance are binned according to the same interval used in the price formula. While the 1-minute system imbalance distribution clearly shifts when more assets start to react to the price signal, the 15-minute distribution remains nearly stable. This difference can be explained by oscillations in the system imbalance on minute-level which cancel out when taking the average over 15-minutes. For the TSO, this observation represents a serious risk, since there is a mismatch between the system imbalances used for imbalance settlement and the actual requirements for grid stability. Nonetheless, we also observe the positive effect of implicit balancing where up to 200 MW of total simulated capacity, the extreme system imbalances are reduced. This also has a positive impact on the balancing costs, as discussed next.

\begin{figure}[h!]
    \centering
    \includegraphics[width=0.86\linewidth]{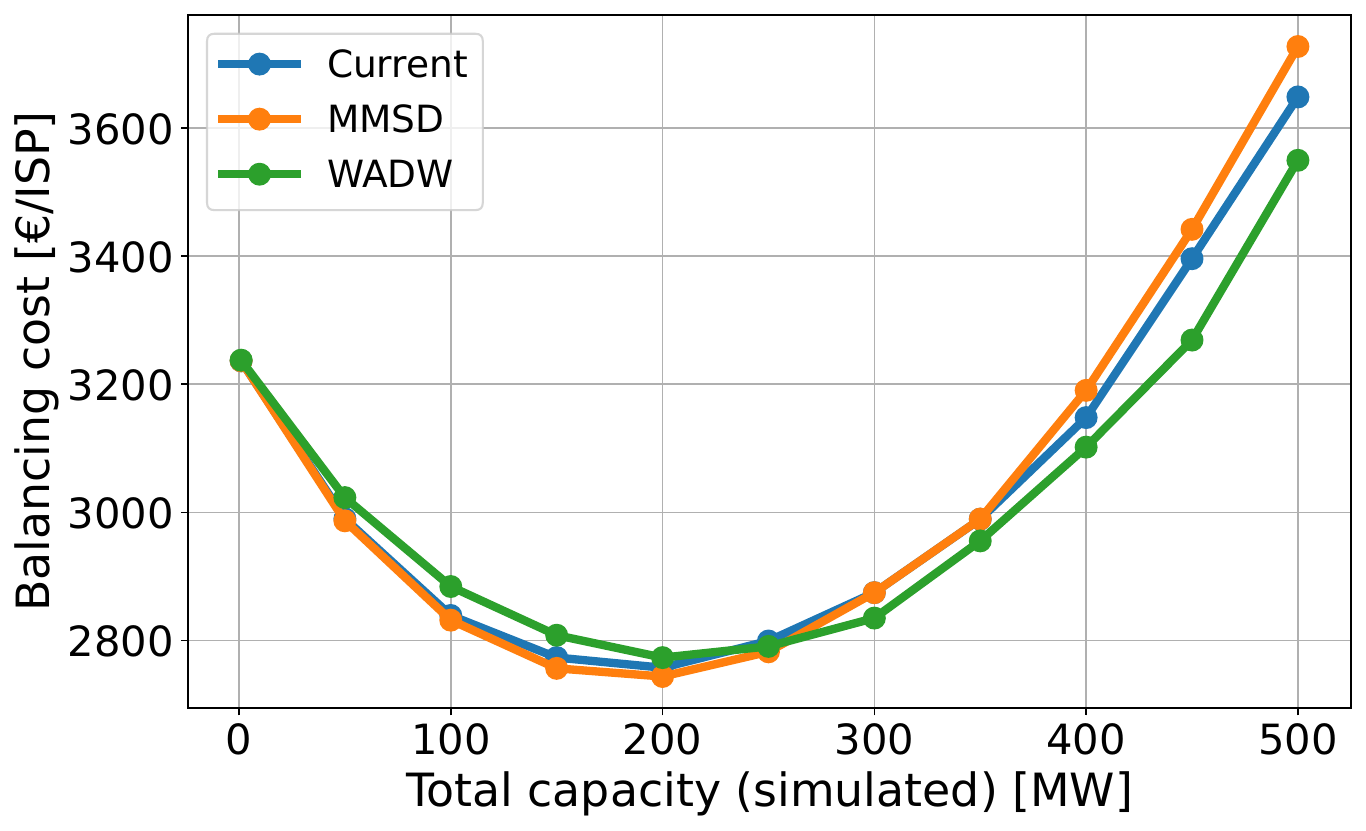}
    \caption{Balancing cost per imbalance settlement period for the different price formulas as a function of the total simulated capacity. The balancing costs include the activation costs of reserves (aFRR/mFRR) and the payment from/to BRPs due to their simulated positions during imbalance settlement.}
    \label{fig:activation_costs}
\end{figure}

Figure \ref{fig:activation_costs} shows how the balancing costs for the TSO evolve when more assets participate in implicit balancing. To this end, we have considered the activation costs of reserves and the payments from/to BRPs for our simulated positions. Note that we do not have information on the actual positions of BRPs and therefore we cannot include the historical payments during imbalance settlement. We observe that, at first, implicit balancing has a positive effect on the total balancing costs for each formula. At 200 MW of additional capacity provided by our simulated assets, the minimum balancing costs are obtained with a decrease of around $12.5\%$. However, when simulating a larger capacity, the balancing costs will start to increase again and from 450 MW on we even observe an increase in total costs. This increase in activation costs is purely driven by the overshoots due to implicit balancing which are covered by explicit activations (and not completely recovered by imbalance settlement). While the general trend between the formulas is similar, we do observe slight differences between the three price formulas. The ``Current'' and ``MMSD'' formulations yield lower activation costs at lower simulated capacities, whereas the ``WADW'' formula attains the lowest cost from 300 MW onward. One possible explanation for this is related to the magnitude of the prices.  The ``Current'' and ``MMSD'' formulas will produce more extreme prices given that they are driven by the min/max operator. Consequently, they will exceed the price thresholds set by the battery controllers more easily and, because of this, trigger more implicit reactions. Initially, this property is advantageous, as the implicit reactions are still contributing to balancing the power grid. However, beyond a certain capacity, this also implies that these two formulas may inadvertently trigger more overshoots compared to the ``WADW'' formula.

\begin{figure}[t!]
    \centering
    \includegraphics[width=0.84\linewidth]{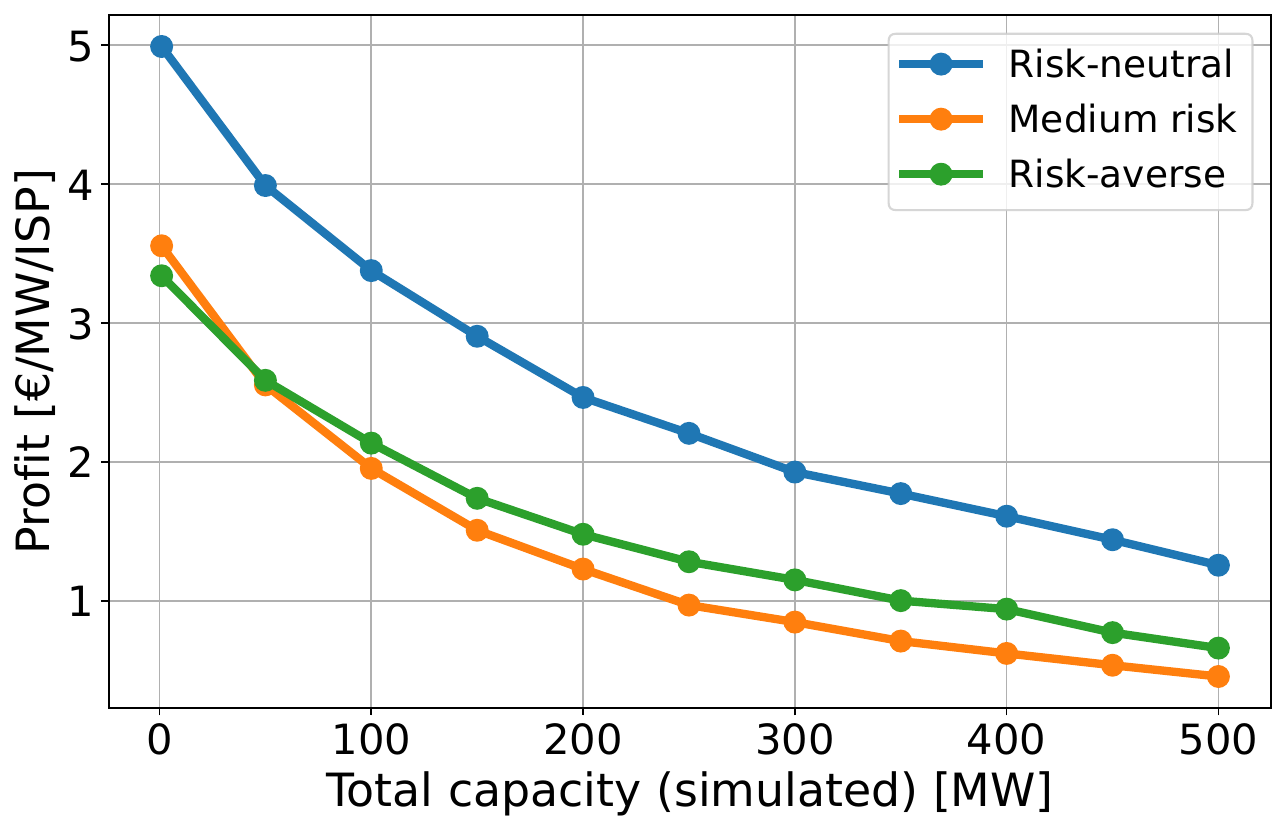}
    \caption{Profit per imbalance settlement period normalized by the power capacity within each risk group. The calculations were made based on the simulations for the current price formula. The profit does not consider any grid transmission fees.}
    \label{fig:bess_profit}
\end{figure}

Figure \ref{fig:bess_profit} shows how the gross profit per imbalance settlement period changes with total simulated capacity. The profit is normalized by the installed capacity to ensure comparability. For each risk group, we observe an initial steep decrease in profit when more assets start to participate, followed by a more gradual decline starting from a capacity of around 200 MW. Nevertheless, even when balancing costs start to increase (at 450 MW or above), it remains profitable for BRPs to pursue implicit balancing with bang-bang control. Additionally, we observe that the risk-averse group achieves higher profitability than the medium-risk group when the total simulated capacity is 100 MW or more. This may be because the risk-averse group has higher thresholds and can therefore only be active in situations where the probability of overshoots is lower.

\section{Discussion and conclusion}
\label{sec:conclusion}

In this work, we studied increasing participation in implicit balancing in the Belgian context. To this end, we developed a market simulator incorporating the feedback loop between the imbalance price and the positions of BRPs (i.e., their implicit response). The market simulator obtained the imbalance price from the observed system imbalance using a mixed integer linear program that models explicit balancing based on the historical individual balancing bids. The imbalance price was computed using the currently adopted formula as well as two future candidate formulas proposed by the Belgian TSO. The implicit response of BRPs was simulated as multiple BESS assets with different risk profiles.

Our findings indicated that while a decentralized balancing model can lower balancing costs, high participation in implicit balancing often triggers overshoots and oscillations when BRPs do not account for their effect on the market. These oscillations represent a risk for the TSO, since they are not properly reflected in the settlement price. Moreover, we found that even though rising balancing costs result in lower profit margins for BRPs, it still remains profitable for them to react to the real-time price signals. Last, the comparison of the different price formulas showed that the ``weighted average dynamic weights'' formula produces less extreme price events and is more robust in case of oscillations of the system imbalance. On the other hand, this formula also provides a less strong signal to the market and therefore might evoke fewer reactions compared to the other two. Still, the three formulas show a similar trend and therefore we believe it is more important to focus on reducing the risk of oscillations during the settlement period.  

Future work could look into ways by which the TSO can reduce this risk. For instance, the TSO can require the assets to react according to a certain shape or apply a penalty to assets which contribute to significant overshoots. Another interesting addition would be to include other asset classes in the simulated response. Wind farms, for example, can curtail their energy production when the system has a surplus, provided that there is sufficient wind at that moment.

\section{Acknowledgments}

The authors gratefully acknowledge the financial support of the Flemish Government through the Flanders AI Research Program and the Sustainable Blue Economy Partnership through the INSPIRE project (project no. SBEP2023-440).

% , the Energy Transition Funds of the Belgian Federal Government through the BeFORECAST project

\FloatBarrier

% \newpage
\printbibliography

@online{Elia2025RealtimePriceII,
  title        = {Public consultation on the “Real-time price” design note {II} focused on the Imbalance Price formula evolutions},
  author       = {{Elia Transmission Belgium SA}},
  year         = {2025},
  month        = {4},
  day          = {4},
  url          = {https://www.elia.be/en/public-consultation/20250404_public-consultation-on-the-real-time-price-design-note-ii},
  urldate      = {2025-09-23},
}

@online{elia2024brp,
  author       = {{Elia Transmission Belgium SA}},
  title        = {{Terms and Conditions for Balance Responsible Parties (BRPs)}},
  year         = {2024},
  month        = {5},
  day          = {22},
  url          = {https://www.elia.be/-/media/project/elia/elia-site/electricity-market-and-system---document-library/balance-responsible-party-and-system-imbalance/2025/tc-brp-en--as-of-010625.pdf},
  urldate      = {2025-09-23},
}

@online{elia2025flexibility,
  author       = {{Elia Transmission Belgium SA}},
  title        = {{Harnessing Flexibility in the Energy Transition}},
  year         = {2025},
  month        = {5},
  url          = {https://issuu.com/eliagroup/docs/harnessing_flexibility_in_the_energy_transistion_?fr=sZjZlOTg0NTQ2NDY},
  urldate      = {2025-10-30},
}

@misc{IEA_GlobalEnergyReview2025,
  author = {IEA},
  title = {Global Energy Review 2025},
  year = {2025},
  publisher = {IEA},
  address = {Paris},
  url = {https://www.iea.org/reports/global-energy-review-2025},
  note = {Licence: CC BY 4.0}
}

@online{acer2020harmon,
  author = {Agency for the Cooperation of Energy Regulators (ACER)},
  title = {Harmonisation of Imbalance Settlement},
    year = {2020},
  url = {https://www.acer.europa.eu/electricity/market-rules/electricity-balancing/harmonisation-of-imbalance-settlement},
  urldate = {2024-10-16}
}

@inproceedings{bruninx2025day,
  title={Day-Ahead Bidding Strategies for Wind Farm Operators Under a One-Price Balancing Scheme},
  author={Bruninx, Max and Verstraeten, Timothy and Kazempour, Jalal and Helsen, Jan},
  booktitle={Proceedings of the 16th ACM International Conference on Future and Sustainable Energy Systems},
  pages={719--726},
  year={2025}
}

@inproceedings{soroush2025gaming,
  title={Gaming strategies in european imbalance settlement mechanisms},
  author={Karimi Madahi, Seyed Soroush and Bruninx, Kenneth and Claessens, Bert and Develder, Chris},
  booktitle={2025 IEEE PES Innovative Smart Grid Technologies Conference Europe (ISGT Europe)},
  pages={1--5},
  year={2025},
  organization={IEEE}
}

@article{smets2023strategic,
  title={Strategic implicit balancing with energy storage systems via stochastic model predictive control},
  author={Smets, Ruben and Bruninx, Kenneth and Bottieau, J{\'e}r{\'e}mie and Toubeau, Jean-Fran{\c{c}}ois and Delarue, Erik},
  journal={IEEE Transactions on Energy Markets, Policy and Regulation},
  volume={1},
  number={4},
  pages={373--385},
  year={2023},
  publisher={IEEE}
}

@article{wessel2024risk,
  title={Risk-aware participation in day-ahead and real-time balancing markets for energy storage systems},
  author={Wessel, Emma and Smets, Ruben and Delarue, Erik},
  journal={Electric Power Systems Research},
  volume={235},
  pages={110741},
  year={2024},
  publisher={Elsevier}
}

@article{rasic2025safe,
  title={Safe Reinforcement Learning for Battery Energy Storage Participation in the Imbalance Settlement},
  author={Rasic, Cyril and Favaro, Pietro and Wang, Yi and Toubeau, Jean-Fran{\c{c}}ois},
  journal={IEEE Transactions on Energy Markets, Policy and Regulation},
  year={2025},
  publisher={IEEE}
}

@article{madahi2025model,
  title={Model Predictive Control-Guided Reinforcement Learning for Implicit Balancing},
  author={Karimi Madahi, Seyed Soroush and Bruninx, Kenneth and Claessens, Bert and Develder, Chris},
  journal={arXiv preprint arXiv:2510.04868},
  year={2025}
}

@article{allard2024forecast,
  title={A forecast-driven stochastic optimization method for proactive activation of manual reserves},
  author={Allard, Julien and Arrigo, Adriano and Bottieau, J{\'e}r{\'e}mie and Bertrand, Gilles and De Gr{\`e}ve, Zacharie and Vall{\'e}e, Fran{\c{c}}ois},
  journal={Electric Power Systems Research},
  volume={235},
  pages={110804},
  year={2024},
  publisher={Elsevier}
}

@online{EliaAdequacyFlexibility2025,
  author      = {{Elia Transmission Belgium SA}},
  title       = {Adequacy and Flexibility Study for Belgium 2026--2036},
  year        = {2025},
  month       = {6},
  day         = {27},
  url         = {https://issuu.com/eliagroup/docs/adequacy_and_flexibility_study_for_belgium_2026-2},
  type        = {Report},
  address     = {Brussels, Belgium}
}

@inproceedings{haaberg2016classification,
  title={Classification of balancing markets based on different activation philosophies: Proactive and reactive designs},
  author={H{\aa}berg, Martin and Doorman, Gerard},
  booktitle={In proceedings of the 13th International Conference on the European Energy Market (EEM)},
  pages={1--5},
  year={2016},
  organization={IEEE}
}

@online{eliaopendata,
  title        = "Elia Open Data Platform",
  author       = {{Elia Transmission Belgium SA}},
  url = {https://opendata.elia.be/},
  urldate={2025-07-01},
}

@online{entsoe_mari,
  author = {{European Network of Transmission System Operators for Electricity (ENTSO-E)}},
  title = {Manual Activation of Reserves Initiative},
  url = {https://www.entsoe.eu/network_codes/eb/mari/},
  urldate = {2025-05-07}
}

@online{entsoe_picasso,
  author = {{European Network of Transmission System Operators for Electricity (ENTSO-E)}},
  title = {{Platform for the International Coordination of Automated Frequency Restoration and Stable System Operation (PICASSO)}},
  url = {https://www.entsoe.eu/network_codes/eb/picasso/},
  urldate = {2025-05-07},
}

\FloatBarrier

\newpage

\appendices
\section{Price calculation}\label{sec:sim-price_components}

In this section, we provide details on the calculation of the imbalance price for the different price components and formulas. With regard to the price components, we only consider the ones currently in place to limit the scope of the experiments. In terms of notation, $T$ is the time of calculation of the imbalance price, whereas $Q$ refers to the imbalance settlement period (which is a set of timesteps $\left\{t_1, \dots, t_{15}\right\}$). $SI$ refers to the cumulative average of the system imbalance over the imbalance settlement period up to timestep T.

% In this section, we provide details on the calculation of the imbalance price for the different price components (Section \ref{sec:ap_price_components}) and formulas (Section \ref{sec:ap_price_formula}).

% The following notation will be used throughout the chapter: $T$ is the time of calculation of the imbalance price, whereas $Q$ refers to the imbalance settlement period (which is a set of timesteps $\left\{t_1, \dots, t_{15}\right\}$). $SI$ refers to the cumulative average of the system imbalance over the imbalance settlement period up to timestep T.

% In this chapter, we provide details on the calculation of the imbalance price for the different formulas. Section \ref{sec:ap_price_components} outlines the different components which are used in the imbalance price formula. While Elia has also outlined different calculations of the price components, we only consider the ones currently in place to limit the scope of the experiments. In Section \ref{sec:ap_price_formula} the different price formulas are discussed.

% The following notation will be used throughout the chapter: $T$ is the time of calculation of the imbalance price, whereas $Q$ refers to the imbalance settlement period (which is a set of timesteps $\left\{t_1, \dots, t_{15}\right\}$). $SI$ refers to the cumulative average of the system imbalance over the imbalance settlement period up to timestep T.

\subsection{Price components}
\label{sec:ap_price_components}

\subsubsection{aFRR component}

The aFRR component is given by the volume weighted average marginal price over all optimization cycles:
\begin{equation}
    \lambda^{\aFRR}=\sum_{t\in Q, t\leq T} \frac{\lambda^{\aFRR^+}_t V^{\aFRR^+}_t + \lambda^{\aFRR^-}_t V^{\aFRR^-}_t}{V^{\aFRR^+}_t + V^{\aFRR^-}_t}.
\end{equation}
% \begin{equation}
%     \lambda^{\aFRR}=\sum_{t\in Q, t\leq T}\frac{\sum_{a^+\in \aFRR^+} \mu^{a^+}b^{a^+}_t + \sum_{a^-\in \aFRR^-} \mu^{a^-}b^{a^-}_t}{\sum_{a^+\in \aFRR^+} b^{a^+}_t + \sum_{a^-\in \aFRR^-} b^{a^-}_t}.
% \end{equation}
Note that in case there are no aFRR activations, this price is set equal to the average of the value of additional activation (VoAA)\footnote{As defined in ``T\&C BRP'' \cite{elia2024brp}: The cheapest energy bid during the imbalance settlement period considering both aFRR and mFRR in the local merit order for a given direction.} in the upward an downward direction. %(and hence, equal to the spot price component defined in (\ref{eq:lambda_spot}))
\\
\subsubsection{mFRR component}

The mFRR component is given by the marginal price of all mFRR activations:
\begin{equation}
    \lambda^{\mFRR} = \begin{cases}
        \max_{t\in Q, t\leq T} \lambda^{\mFRR^{+}}_t\quad SI\leq0\\
        \min_{t\in Q, t\leq T} \lambda^{\mFRR^{-}}_t\quad SI >0.
    \end{cases}
\end{equation}
Note that the mFRR component does not exist in case there are no mFRR activations.
\\
\subsubsection{Spot price component}

The spot price component is given by average between the value of additional activation (VoAA) in the upward and downward direction:
\begin{equation}
    \lambda^{\mathrm{spot}} = \frac{\text{VoAA}_+ + \text{VoAA}_-}{2}.\label{eq:lambda_spot}
\end{equation}
Note that in the current price calculation Elia refers to this component as the deadband value.
\\
\subsubsection{Alpha component}
\label{sec:sim-price_components_alpha}

The alpha component is a correction to low imbalance prices for large system imbalances (with an absolute value larger than 150 MW) and is given by:
\begin{equation}
    \alpha = \left(a + \frac{b}{1+e^\frac{c-x}{d}}\right) cp,
\end{equation}
with $a=0,\ b=200,\ c=450,\ d=65,\ x=$ average between the system imbalance during the current ISP and the system imbalance during the previous ISP and $cp$ defined as:
\begin{align}
    cp =\begin{cases}
        \mathrm{clip}(\frac{400-\lambda^{\B}}{200})& SI<-150,\\
        -\mathrm{clip}(\frac{\lambda^{\B}+200}{200})& SI>150\\
        0& SI \in [-150, 150],
    \end{cases}
\end{align}
with $\mathrm{clip}(x) = \min\left(1, \max(0,x)\right)$ a function which clips the values between 0 and 1.\\
\\
After obtaining this value, we update the balancing price accordingly:
\begin{equation}
    \lambda^{\B}\leftarrow\lambda^{\B}+\alpha.
\end{equation}

\vspace{8pt}

\subsection{Price formulas}
\label{sec:ap_price_formula}

\vspace{12pt}

\subsubsection{Current formula}
The current balancing price formula (as described in \cite{elia2024brp}) is given by:

% \begin{subequations*}
\begin{align}
    \lambda^{\B} = \begin{cases}
        \lambda^{\mathrm{spot}}& SI\in[-25,25]\\
        \max{(\lambda^{\aFRR}, \lambda^{\mFRR})}& SI < -25\\
        \min{(\lambda^{\aFRR}, \lambda^{\mFRR})}& SI > 25.
    \end{cases}\label{eq:sim-current_price}
\end{align}
% \end{subequations*}
When the system imbalance is small, the balancing price should not incentivize BRPs to react and therefore it is set equal to the spot price component between -25 MW and 25 MW. Note that here we exclude the floor and the cap components since they are introduced to mitigate certain effects related to European balancing platforms (PICASSO/MAARI).\\
\\
\subsubsection{Max/min with smoothed deadband}
In the first proposition by Elia to calculate the balancing price, a continuous weight (as a function of system imbalance) is attributed to the spot price component rather than a fixed deadband:

\begin{equation}
    \lambda^{\B} = w^{\textrm{spot}}\:\lambda^{\mathrm{spot}} + (1-w^{\textrm{spot}})\:\lambda^{\mathrm{FRR}},\label{eq:sim-prop_I}
\end{equation}
with:
\begin{equation}
    w^{\textrm{spot}} = e^{-(\frac{SI}{25})^4},\label{eq:sim-weight_fun}
\end{equation}
\begin{equation}
    \lambda^{\mathrm{FRR}} = \begin{cases}
        \max{(\lambda^{\aFRR}, \lambda^{\mFRR})}\quad SI \leq 0\\
        \min{(\lambda^{\aFRR}, \lambda^{\mFRR})}\quad SI > 0.
    \end{cases}
\end{equation}

\noindent The advantage of this approach is that it results in a smooth transition between the spot price component and the aFRR/mFRR component rather than jumps at the boundary values. It should be noted that we slightly modified the weight function (\ref{eq:sim-weight_fun}) compared to the example function given by Elia to more closely resemble the boundaries in (\ref{eq:sim-current_price}).\\
\\
\subsubsection{Weighted average with dynamic weights}
In the second proposition, the balancing price consists of the weighted average of the aFRR and mFRR price components rather than a max/min approach.

\begin{align}
    \lambda^{\B} = 
        w^{\aFRR} \lambda^{\aFRR} + (1-w^{\aFRR})\lambda^{\mFRR},
\label{eq:sim-prop_II}
\end{align}
with:
\begin{equation}
    w^{\aFRR} = \sum_{t\in Q, t\leq T}\frac{\sum_{a^+\in \aFRR^+} b^{a^+}_t+ \sum_{a^-\in \aFRR^-} b^{a^-}_t}{\sum_{r\in\mathcal{R}}b^r_t}.
\end{equation}

\FloatBarrier

\section{Sensitivity study}
\label{sec:app_sensitivity_study}
\raggedbottom

In this section we provide the same figures shown in Section \ref{sec:results} for two different distributions of the total power among the risk groups: (1) 60\% risk-neutral, 20\% medium-risk, 20\% risk-averse (Section \ref{sec:app_alt_1}) and (2) 20\% risk-neutral, 20\% medium-risk, 60\% risk-averse (Section \ref{sec:app_alt_2}). Note that here we only consider the current price formula in order to reduce the computation time of the experiments.

\FloatBarrier

\subsection{Majority = risk-neutral}
\label{sec:app_alt_1}

\begin{figure}[H]
    \centering
    \includegraphics[width=0.8\linewidth]{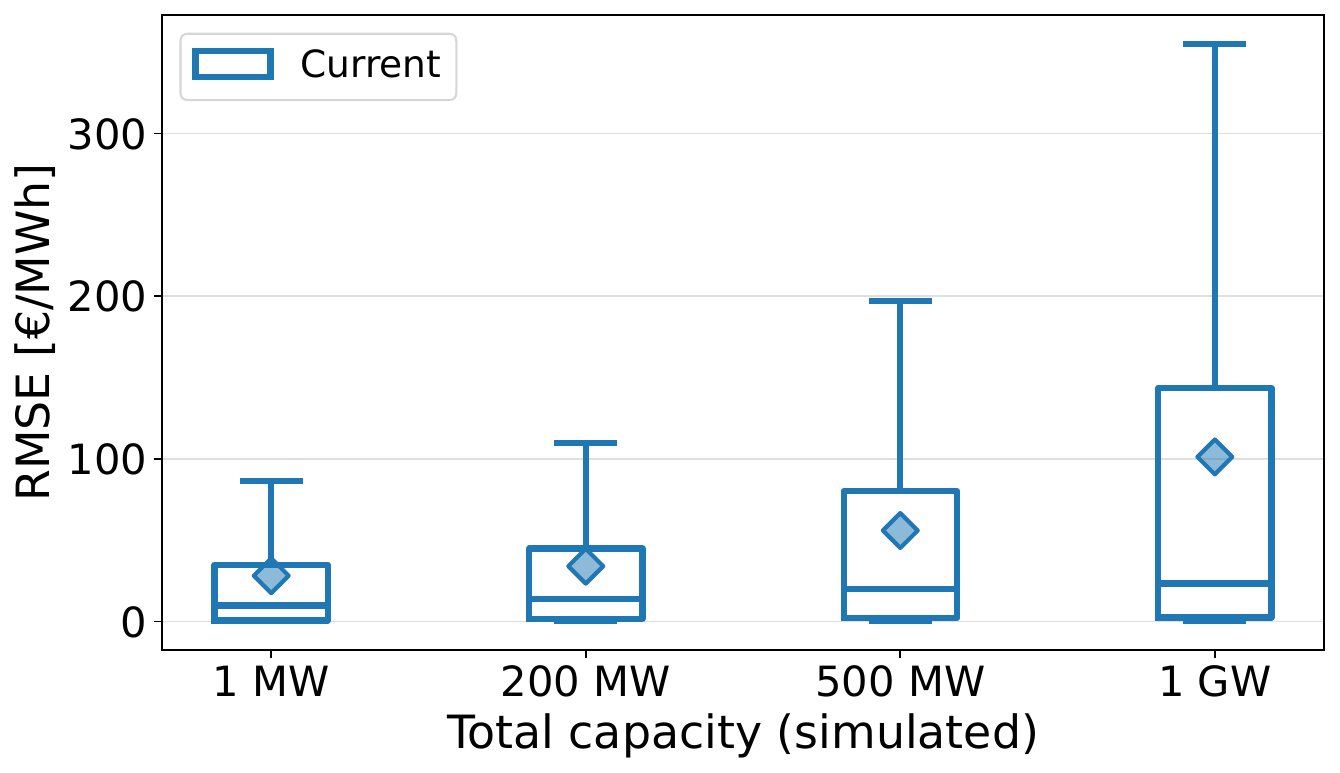}
    \caption{Boxplots of the RMSE of the minute-level imbalance price compared to the settlement price, with the diamond marker indicating the average value. The RMSE is calculated for each imbalance settlement period seperately.}
    % \label{fig:rmse}
\end{figure}

\begin{figure}[H]
    \centering
    \includegraphics[width=0.8\linewidth]{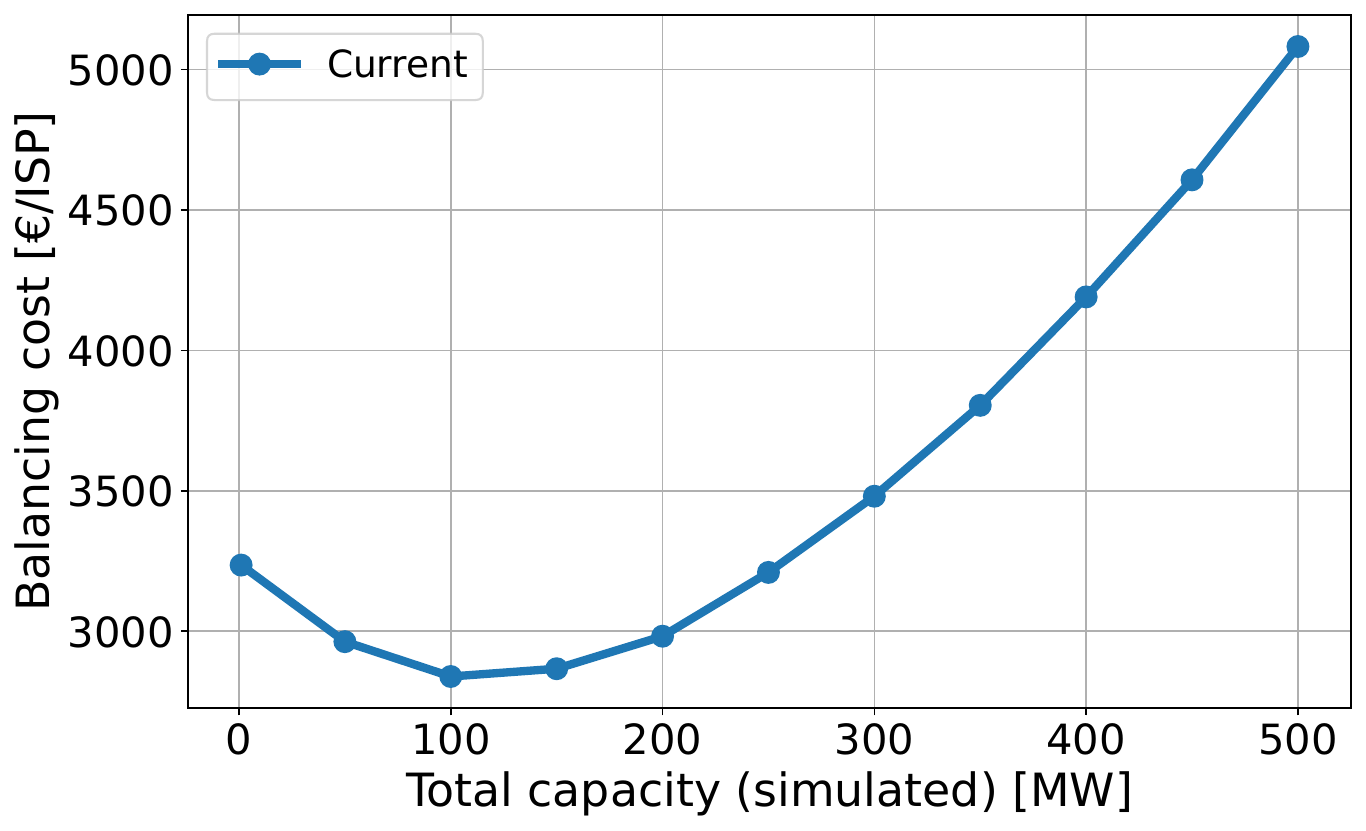}
    \caption{Balancing cost per imbalance settlement period as a function of the total simulated capacity. The balancing costs include the activation costs of reserves (aFRR/mFRR) and the payment from/to BRPs due to their simulated positions during imbalance settlement.}
    % \label{fig:activation_costs}
\end{figure}

\begin{figure}[H]
    \centering
    \includegraphics[width=0.8\linewidth]{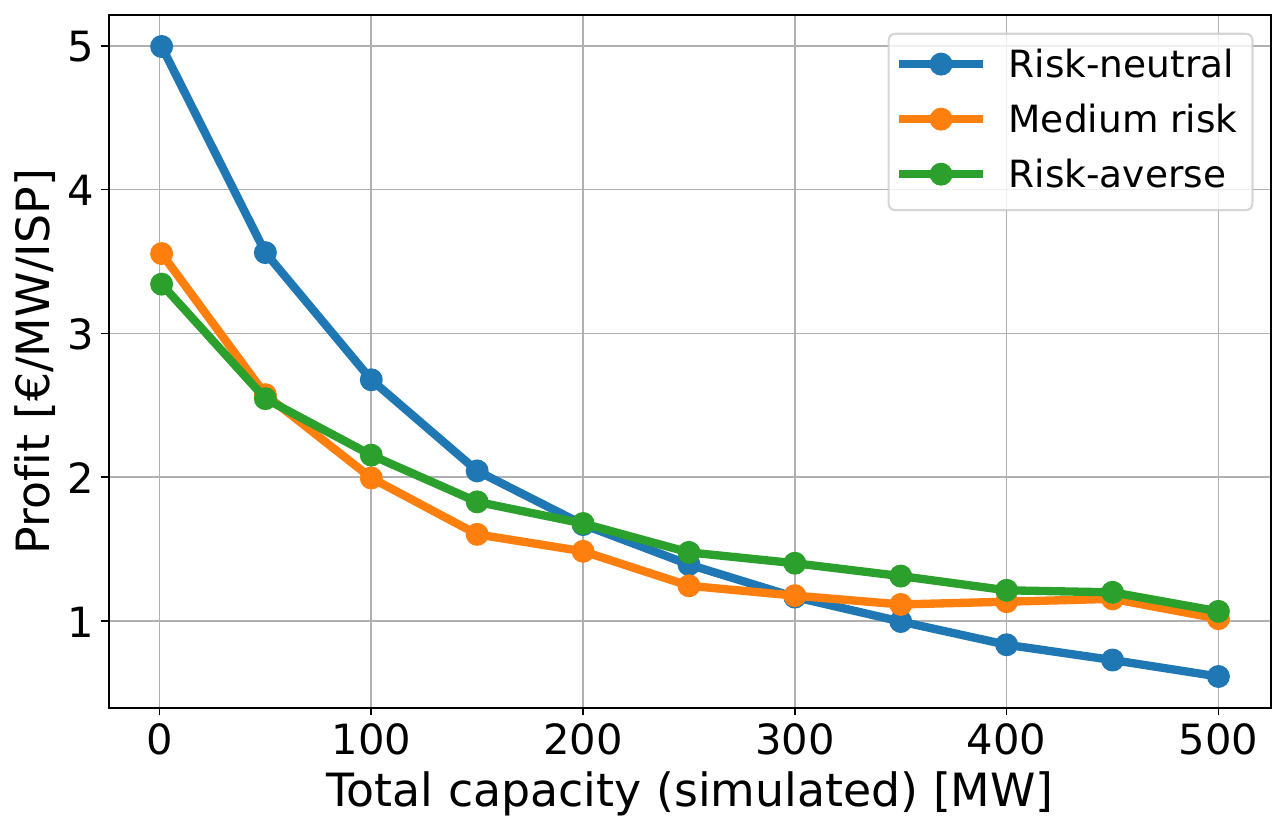}
    \caption{Profit per imbalance settlement period normalized by the power capacity within each risk group.}
    % \label{fig:bess_profit}
\end{figure}

\begin{figure}[H]
     \centering
     \begin{subfigure}[b]{0.4\textwidth}
         \centering
         \includegraphics[width=\textwidth]{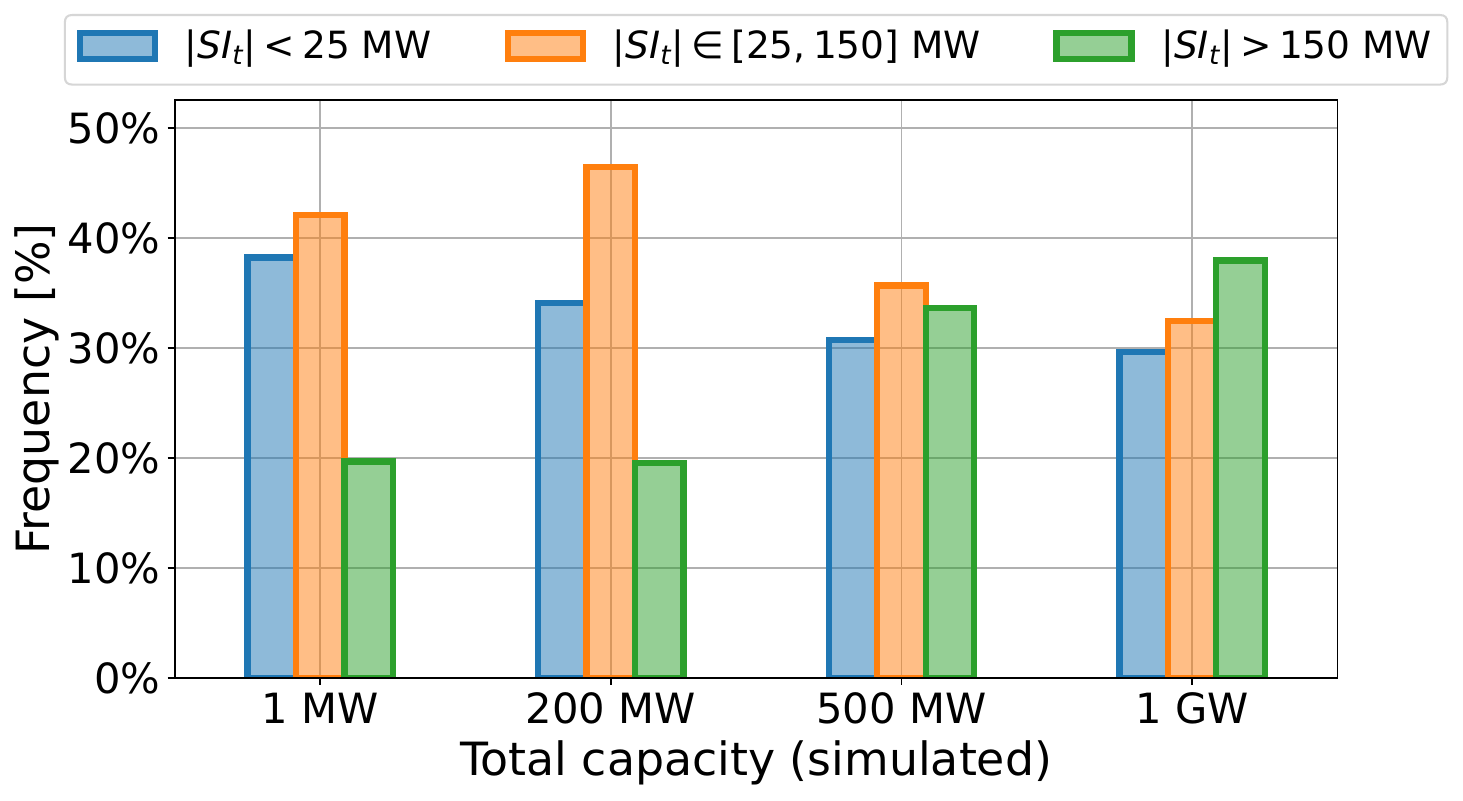}
         \caption{1-min}
         % \label{fig:system_imb_1min}
     \end{subfigure} 
     \vspace{12pt}
     \begin{subfigure}[b]{0.4\textwidth}
         \centering
         \includegraphics[width=\textwidth]{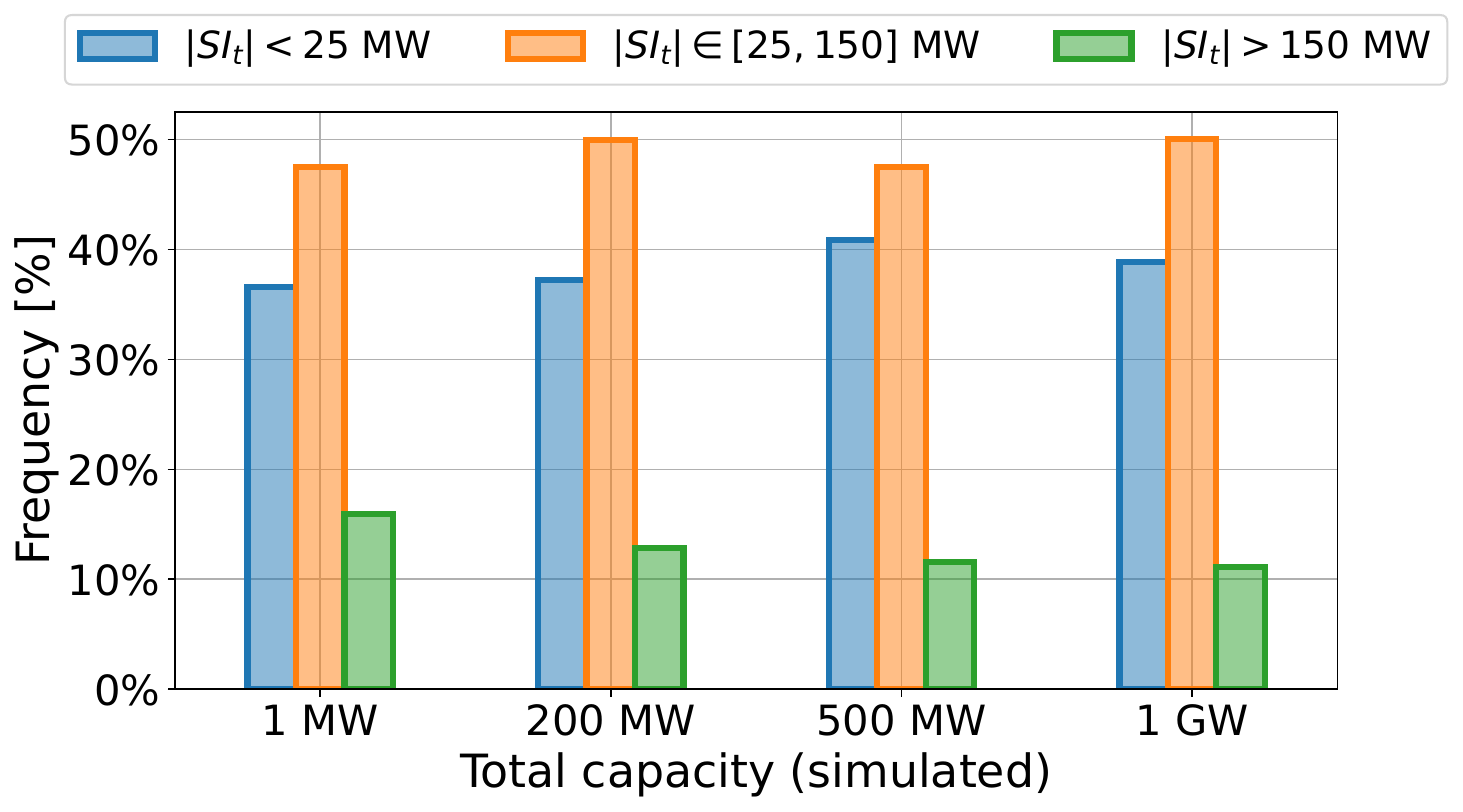}
         \caption{15-min}
         % \label{fig:system_imb_15min}
     \end{subfigure}
     \caption{Distribution of the absolute system imbalance (1-minute and 15-minute level). The imbalances are binned according to the three regions considered in the price formula: the deadband ($<$$25$  MW), the imbalance tariff without alpha factor ($25$$-$$150$ MW), and the imbalance tariff with alpha factor ($>$$150$ MW).}
     % \label{fig:system_imbalance}
\end{figure}

\FloatBarrier

\subsection{Majority = risk-averse}
\label{sec:app_alt_2}

\begin{figure}[H]
    \centering
    \includegraphics[width=0.8\linewidth]{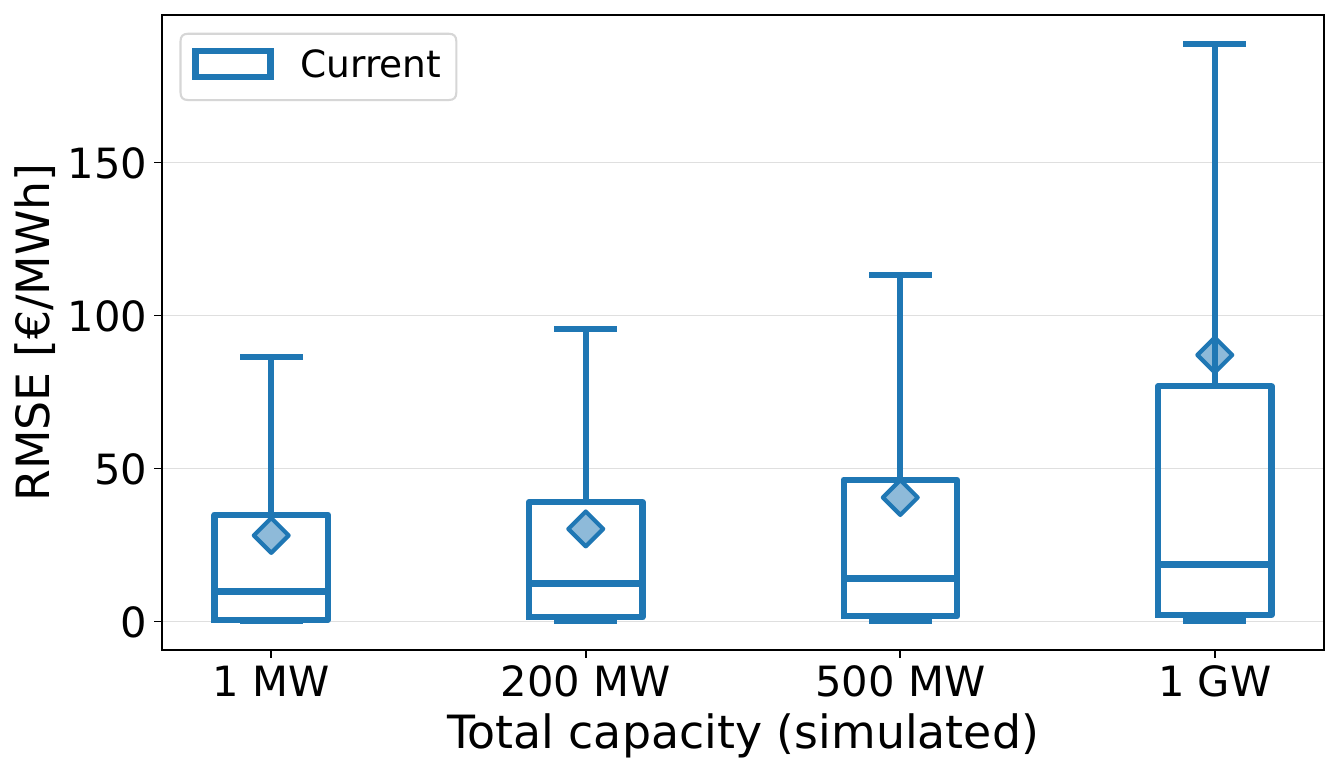}
    \caption{Boxplots of the RMSE of the minute-level imbalance price compared to the settlement price, with the diamond marker indicating the average value. The RMSE is calculated for each imbalance settlement period seperately.}
    % \label{fig:rmse}
\end{figure}

\begin{figure}[H]
    \centering
    \includegraphics[width=0.8\linewidth]{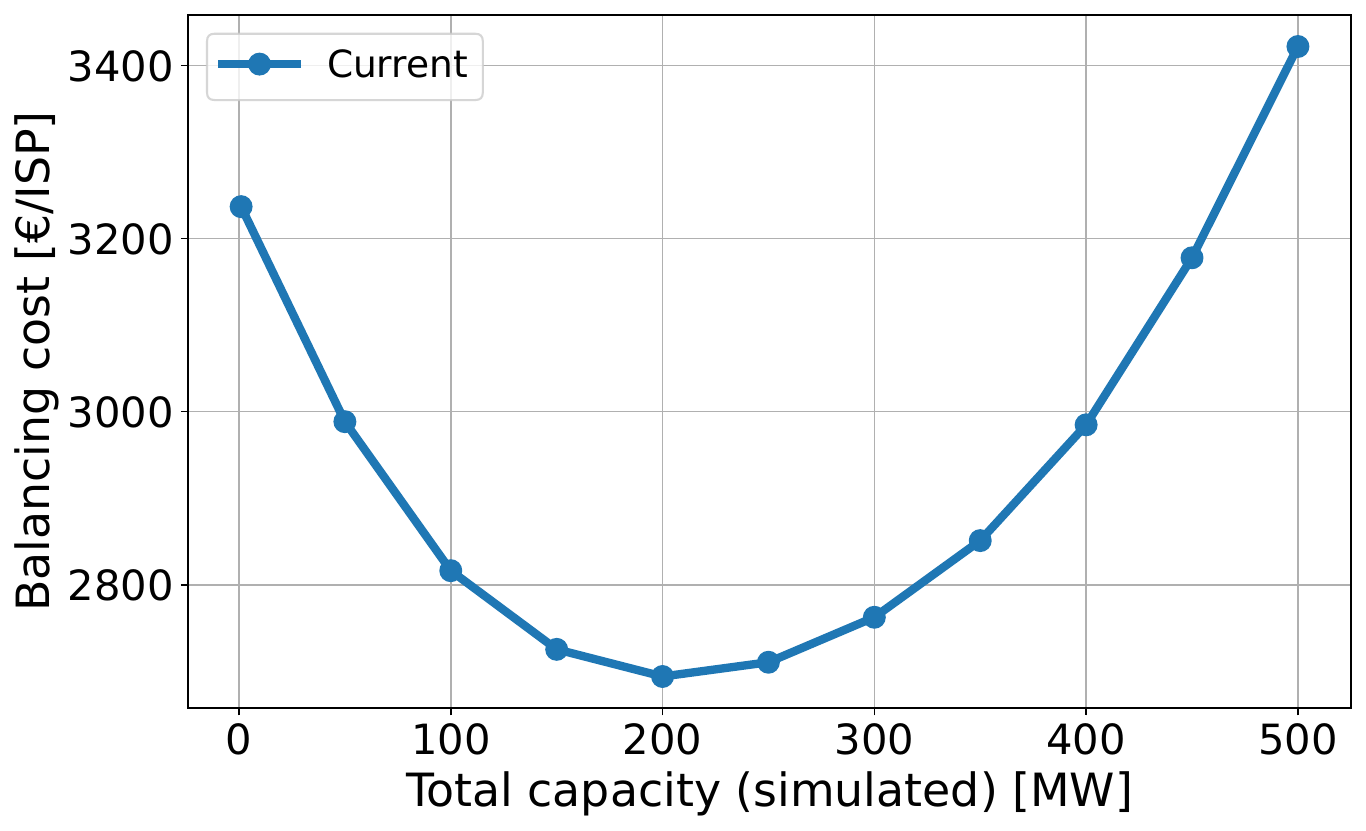}
    \caption{Balancing cost per imbalance settlement period as a function of the total simulated capacity. The balancing costs include the activation costs of reserves (aFRR/mFRR) and the payment from/to BRPs due to their simulated positions during imbalance settlement.}
    % \label{fig:activation_costs}
\end{figure}

\begin{figure}[H]
    \centering
    \includegraphics[width=0.8\linewidth]{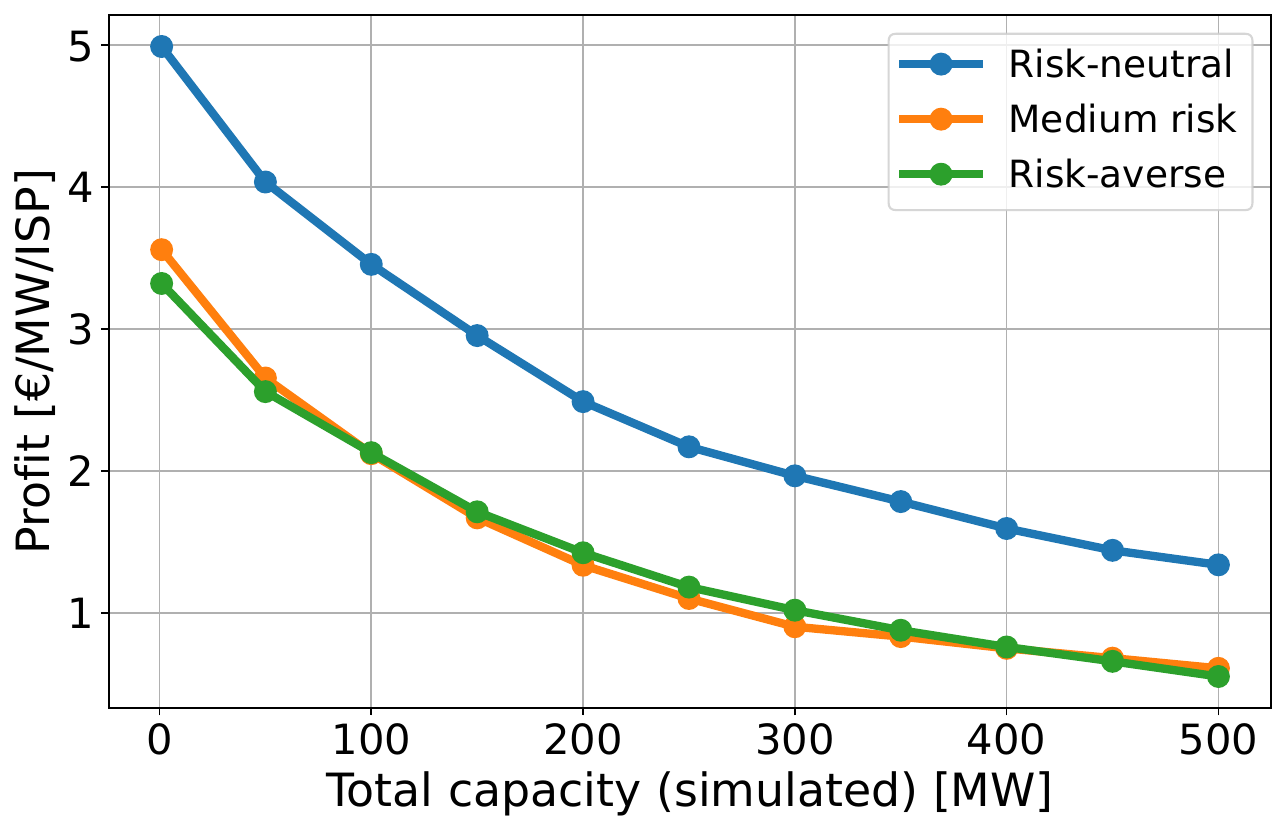}
    \caption{Profit per imbalance settlement period normalized by the power capacity within each risk group.}
    % \label{fig:bess_profit}
\end{figure}

\begin{figure}[H]
     \centering
     \begin{subfigure}[b]{0.4\textwidth}
         \centering
         \includegraphics[width=\textwidth]{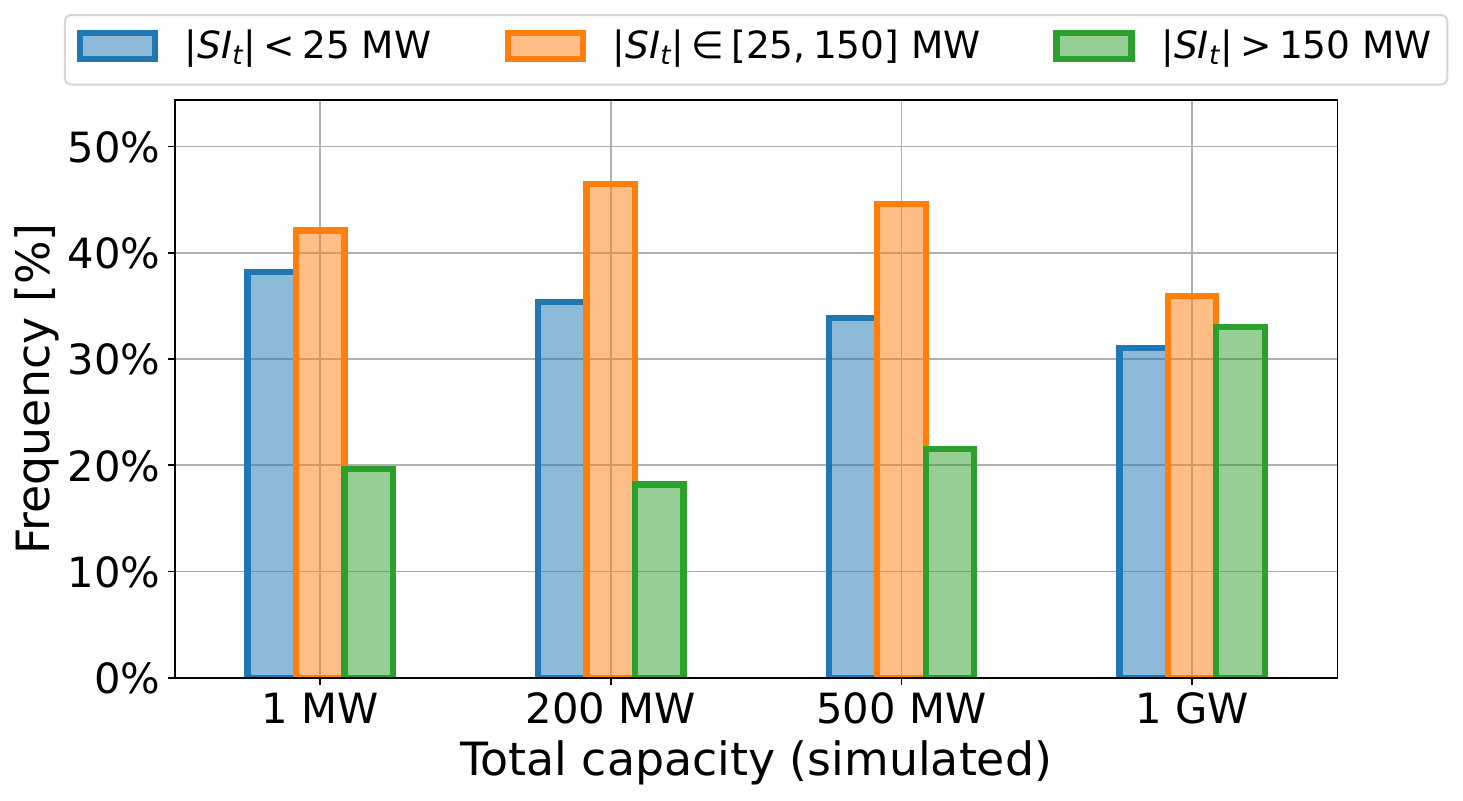}
         \caption{1-min}
         % \label{fig:system_imb_1min}
     \end{subfigure}
     
     \vspace{12pt}
     
     \begin{subfigure}[b]{0.4\textwidth}
         \centering
         \includegraphics[width=\textwidth]{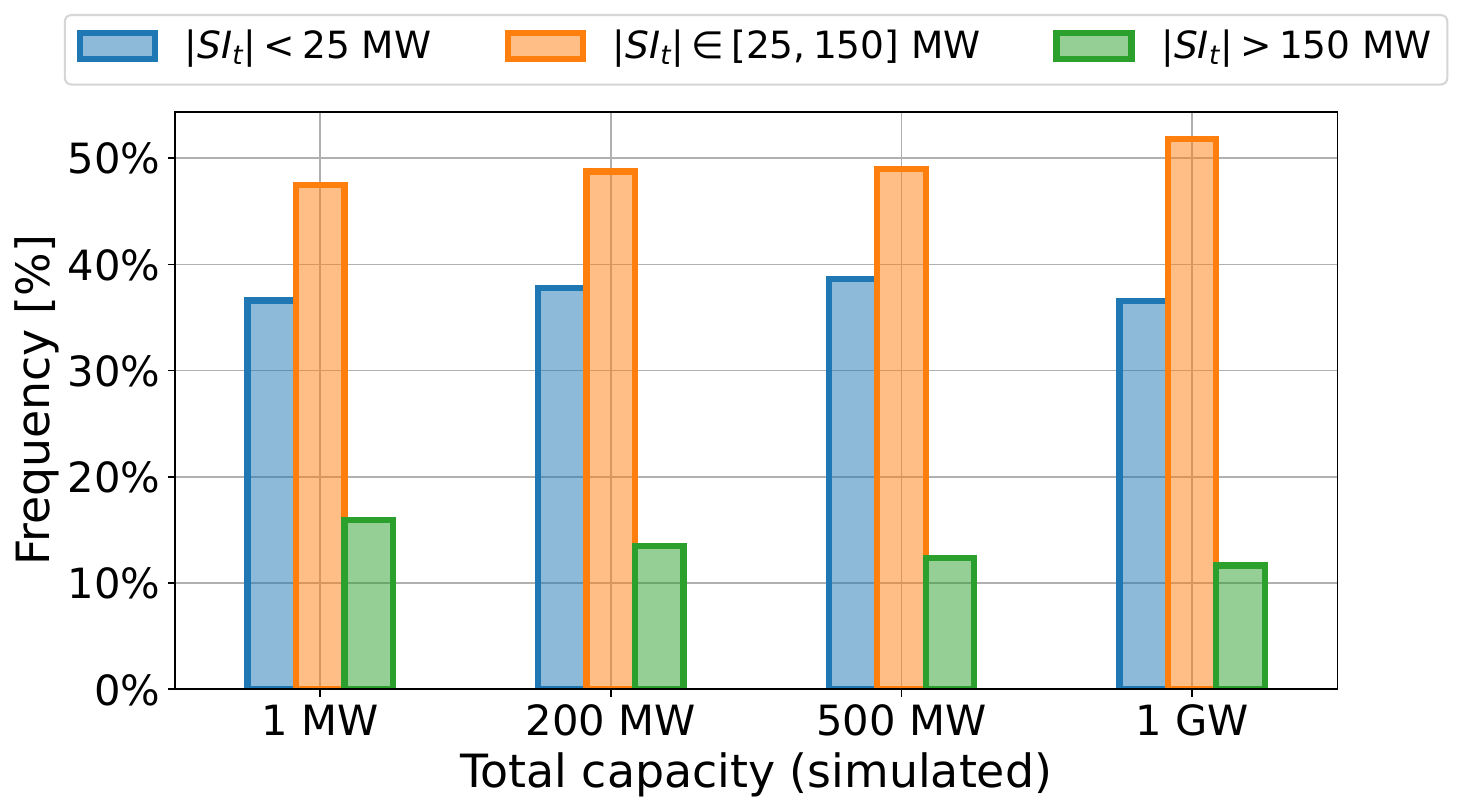}
         \caption{15-min}
         % \label{fig:system_imb_15min}
     \end{subfigure}
     \caption{Distribution of the absolute system imbalance (1-minute and 15-minute level). The imbalances are binned according to the three regions considered in the price formula: the deadband ($<$$25$  MW), the imbalance tariff without alpha factor ($25$$-$$150$ MW), and the imbalance tariff with alpha factor ($>$$150$ MW).}
     % \label{fig:system_imbalance}
\end{figure}

\end{document}